\documentclass{ifacconf}

\usepackage{amsfonts}
\usepackage{amsmath}
\usepackage{amssymb}
\usepackage{color}
\usepackage[numbers]{natbib}
\usepackage{graphicx}
\usepackage{listings}
\usepackage{xspace}
\usepackage{tikz}
\usepackage{subcaption}
\usetikzlibrary{calc, positioning, fit, arrows, arrows.meta, shapes.geometric, shapes.symbols, shapes.misc, patterns}

\newcommand{\Zeta}{Z}

\begin{document}
\begin{frontmatter}

\title{The Countries' Relation Formation Problem: I and II} 

\thanks[footnoteinfo]{This work was supported by National Science Foundation grant n.1607101.00 and US Air Force grant n. FA9550-16-1-0290. Proofs of Theorem 1, Propositions 4 and 5 will appear elsewhere.}
\thanks[footnoteinfo]{Y. Li and A. S. Morse are respectively with the department of Political Science and the department of Electrical Engineering at Yale University, New Haven, CT, US, \{yuke.li, as.morse\}@yale.edu.}

\author{Yuke Li and A. Stephen Morse} 


%
%

\begin{abstract}This paper integrates the studies of various countries' behaviors, e.g., waging wars and entering into military alliances, into a general framework of \emph{countries' relation formation}, which consists of two components,  i.e., a static game and a dynamical system. Aside from being a stand-alone framework itself, this paper can also be seen as a necessary extension of a recently developed \emph{countries' power allocation game} in  \cite{allocation}. We establish certain theoretical results, such as pure strategy Nash equilibrium existence in the static game, and propose several applications of interest made possible by combining both frameworks of countries' power allocation and relation formation. 


\end{abstract}


\end{frontmatter}

\section{Introduction}
\label{intro}

Countries pursue changes in their relations all of the time. A country's entering into a military alliance or any form of cooperative ventures with other countries can be regarded as its formation of  \text{friend} relations with them; A country's making threats to use force, escalating a militarized conflict, or declaring a war on other countries, can be regarded as its formation of \text{adversary} relations with them. Even a country's cancellation of these said relations with other countries, e.g., withdrawing from a military alliance, can be regarded as its formation of a ``\text{null}'' relation with them.  

We summarize the above behaviors into \emph{countries' relation formation}.  With another, a country can form:
\begin{enumerate}
\item   a ``\text{friend}'' relation, which ranges from formal military alliances to informal commitments;

\item an ``\text{adversary}'' relation, which ranges from wars to threats of use of force;
\item a ``\text{null}'' relation, which implies no specific cooperative or conflictual elements.  As mentioned, a \text{null} relation could be formed through canceling an existing \text{friend} relation (e.g., canceling the alignment) or \text{adversary} relation (e.g., having a ceasefire). 

\end{enumerate}

Motivated by the fact that a country may make the joint decisions about choosing its allies and adversaries, such as Germany signing the Molotov-Ribbentrop pact with the Soviet Union shortly before invading Poland, we make another abstraction that a country simultaneously forms different types of relations with different countries. 

We thus unify the studies of multiple types of countries' relation formation behaviors in a single theoretical framework. The existing literature usually focuses on either countries' formation of \text{friend} relations (i.e., alliance formation) by taking the \text{adversary} relations among them (i.e., wars) as given (e.g., \cite{smith1995alliance}) or their formation of \text{adversary} relations (i.e., causes of wars)  by taking \text{friend} relations among them (i.e., alliances) as given (e.g., \cite{leeds2003alliances}).

Adapted from Chapter 3 of \cite{yukedis}, this framework consists of two models, a static game and a dynamical system, which share the same setup, i.e., the same set of assumptions about countries (e.g., their strategies of forming relations, and the realistic rules governing relation formation), discussed in Section~\ref{strategy}. Section~\ref{sg} and Section~\ref{deviation} follow up by respectively developing the two models. 

Next we argue in Section~\ref{linkage} why the framework of countries' relation formation is also a necessary extension of countries' power allocation game in networked international environments in \cite{allocation}. Intuitively, countries change the relations to change the international environments toward where they can more easily pursue their targets with power allocation. Therefore, the outcomes from the power allocation games in different environments thus provide a valid rationale for assessing the extent to which countries will prefer the new environment after changing their relations. And we formalize this rationale using a set of ``basic choice axioms'' that should be innate to any country's preferences for different ``relation configurations'' (which consist of the relations of all countries) formed as a result of their relation formation behaviors. 

 Lastly, we discuss two topics --- \emph{alliance politics} and \emph{great powers' optimal network design} --- as direct applications of the combination of the two frameworks of countries' power allocation and relation formation in Section~\ref{application}.

\section{Setup}
\label{strategy}

As in \cite{allocation},  there is a collection $\mathcal{C}$ of $n$ countries, labeled $1,2, ..., n$; let the set of labels be $\mathbf{n} = \{1,2,3, ..., n\}$.  Any country $i \in \mathbf{n}$ has a \emph{linking strategy}, which is to choose a relation out of the three possibilities, a \emph{\text{friend}} relation, an \emph{\text{adversary}} relation or a \emph{\text{null}} relation, to be formed with every other country.  A vector $L_{i} = [l_{ij}]^{1 \times n} \in \mathcal{L}_{i}$, describes its linking strategy, where

\begin{enumerate} 

\item $l_{ij} = \text{friend}$ or $\text{adversary}$ means country $i$ ``links'' with $j$ to establish a friend relation or an \text{adversary} relation. 
\item If $l_{ij} = \text{null}$ means country $i$ ``unlinks'' with $j$, which can also be regarded a special form of linking.
\item $l_{ii} = \text{self}$ by default.

\end{enumerate} 

Country $i$'s \emph{linking strategy set} is $\mathcal{L}_{i}$, whose cardinality is $3^{n-1}$. The Cartesian product of all countries' linking strategy sets is the \emph{linking matrices set}, $\mathcal{L} = \prod_{i=1}^n \mathcal{L}_{i}$, whose cardinality is $|\mathcal{L}| = 3^{n(n-1)}$.

By a \emph{relation configuration} we mean a configuration of the relations among all countries in $\mathbf{n}$. A relation configuration is represented by a symmetric matrix $R = [r_{ij}]^{n \times n}$, where $r_{ii} = \text{self}$, $r_{ij} \in \{\text{friend},\text{adversary},\text{null}\}$, and $r_{ij} = r_{ji}$. The \emph{relation configurations set} is denoted as $\mathcal{R}$, whose cardinality is $3^{n(n-1)/2}$. 



We now provide below the common-sense rules of how two countries' linking strategies give rise to one of the three relations between them. For instance, bilateral linking by the two countries toward each other is both necessary and sufficient for them to form a friend relation; however, unilateral linking by one of the two countries toward the other is already sufficient for forming an adversary relation between them.  Formally, a relation function $\tau: \mathcal{L} \to \mathcal{R}$ maps a linking matrix $L \in \mathcal{L}$ to a relation configuration $R \in \mathcal{R}$ based on the following assumptions:

\begin{enumerate}
\item  $r_{ij} = r_{ji} = \text{friend}$ if and only if $l_{ij} = l_{ji} = \text{friend}$. \\
\item  $r_{ij} = r_{ji} = \text{adversary}$ if and only if $l_{ij} = \text{adversary}$, or $l_{ji} = \text{adversary}$ or both. \\
\item  $r_{ij} = r_{ji} = \text{null}$ if and only if $l_{ij} = \text{null}$ and $l_{ji} \neq \text{adversary}$, or $l_{ji} = \text{null}$ and $l_{ij} \neq \text{adversary}$ or both. \\
\end{enumerate}

Any country $i \in \mathbf{n}$ ranks over all possible relation configurations in $\mathcal{R}$ using a valid total order. We define this ranking as its \emph{preferences for relation configurations}, written formally as $(\mathcal{R}, \preccurlyeq)$. For country $i$, given any $R$, $\hat{R} \in \mathcal{R}$,  $R \preccurlyeq \hat{R}$  means that country $i$ weakly prefers $\hat{R}$ over $R$, i.e., strictly preferring $\hat{R}$ over $R$  ($R \prec \hat{R}$) or being indifferent between them ($R \sim \hat{R}$).


\section{The Theoretical Framework}

\subsection{Static Game}
\label{sg}

In the static game of interest, any country $i$ simultaneously and optimally picks its linking strategy $L_{i}$  in order to form relations with other countries based on its preferences for the relation configurations that form, $(\mathcal{R},\preccurlyeq)$.   The game can be denoted as the collection of the aforementioned parameters, i.e., $\Sigma = \{\mathcal{C}, \mathcal{L}, \tau, \mathcal{R},\preccurlyeq\}$.   A \emph{complete information} framework is also assumed, where countries have full knowledge of the elements in $\{\mathcal{C}, \mathcal{L}, \tau, \mathcal{R},\preccurlyeq\}$. 
 
An equilibrium relation configuration emerges if and only if every country's linking strategy is in best response to one another, i.e., no one strictly benefits by changing its own linking strategy.   A linking matrix or (a linking strategy profile) $L$ is a Nash equilibrium if for each country $i \in \mathbf{n}$ and for any matrix $\hat{L} \in \mathcal{L}$ in which the $i$-th row of $\hat{L}$ is different from that of $L$, $$\tau(\hat{L}) \preccurlyeq \tau(L) ~\text{or equivalently}~ \hat{R} \preccurlyeq R$$

%

\begin{prop}
\label{prop/r0}
The game $\Sigma = \{\mathcal{C}, \mathcal{L}, \tau, \mathcal{R},\preccurlyeq\}$ always has a pure strategy Nash equilibrium $L^{*}$, where for $i$ and $j \in \mathbf{n}$, $l_{ij} = l_{ji} = \text{adversary}$. 
\end{prop}

\noindent {\bf Proof of Proposition 1:} Let $L$ be such that for any two countries in $\mathbf{n}$, $i$ and $j$,
\begin{equation*}
l_{ij} = l_{ji} = \text{adversary}.
\end{equation*}

No country can unilaterally change any of its \text{adversary} relations. Therefore, no one will deviate, and $L$ is by default a pure strategy Nash equilibrium. $\square$

\subsection{Dynamical System: Distributed Deviation} 
\label{deviation}

In this section we provide an opposite view to the static game, which is a dynamical system, to study countries' relation formation behaviors as a \emph{distributed, deviation process}.

\subsubsection{Countries' Deviations} At time $t+1$, each country updates its own linking strategy based on the strategies of others at time $t$. While updating its own strategy, each country $i \in \mathbf{n}$ implicitly assumes that the strategies of all other countries remain the same. 

For simplicity, let the linking matrix at time $t$ be $L_{u}$ and the linking matrix at time $t+1$ be $L_{v}$, where $L_{u}, L_{v} \in \mathcal{L}$.

\subsubsection{Set of Deviations from A Linking Matrix}  

From $L_{u}$, country $i$ can only deviate to a subset of $\mathcal{L}$, which we define as \emph{$i$'s set of deviations from $L_{u}$} and denote using $\mathcal{L}^{i}_{u}$, of which $L_{u}$ is by default an element. In $\mathcal{L}^{i}_{u}$, a linking matrix that is not $L_{u}$ is only different from $L_{u}$ in terms of the $i$-th row. 


\subsubsection{Individual Transition Probability} Any country chooses a time-invariant nonnegative probability called \emph{individual transition probability} with which to transition from a linking matrix to another. Formally speaking, the map 
\begin{equation*}
\psi_{i}: \mathcal{L} \times \mathcal{L} \to [0,1]
\end{equation*}
denotes country $i$'s transition probability from a linking matrix to another.

If $L_{v}$ is not in $i$'s deviation set from $L_{u}$, we stipulate that $\psi_{i}(u,v)  = 0$; otherwise, $\psi_{i}(u,v) $ is a time-invariant function of country $i$'s preferences for the relation configurations specified in the game $\Sigma$.

\subsubsection{Linking Matrix Transition Probability}

Let the transition probability matrix of the linking matrices be $\Zeta = [\zeta_{uv}] \in [0,1]^{|\mathcal{L} | \times |\mathcal{L}|} =  [\zeta_{uv}] \in [0,1]^{|3^{n(n-1)}| \times |3^{n(n-1)}|}$, where $\zeta_{uv}$ represents the \emph{transition probability} from a linking matrix $u$ to another $v$.  In this paper we focus on time-invariant $\Zeta$.

\subsubsection{Deviation Order} 

 We turn to discuss how countries' deviation order (i.e., asynchronous or synchronous) impacts the way in which the individual transition probability from $L_{u}$ to $L_{v}$, $\psi_{i}(L_{u},L_{v})$, aggregates to the linking matrix transition probability from $L_{u}$ to $L_{v}$, $\zeta_{uv}$.  Two possible deviation orders are:

\begin{enumerate}

\item \emph{Asynchronous Updating.} The first deviation rule is where only one country deviates at each time.

\begin{enumerate}
\item {\it Received Chance of Deviation.} We suppose that only one country is picked based on a nonnegative probability $a_{i}$ to deviate at any time instant, where the total probabilities for all countries sum to 1 as below.  The received chance for country $i$ to deviate can be a good proxy of its relative capabilities in shaping the international environments. 

\begin{equation*}
a_{i} \in [0,1], ~\texttt{and}~ \sum\limits_{i\in \mathbf{n}}a_{i} = 1
\end{equation*}

\item {\it Linking Matrix Transition Probability.} The transition probability from $L_{u}$ to $L_{v}$, $\zeta_{uv}$, can be expressed as 
\begin{equation*}
\zeta_{uv} = \sum\limits_{i\in \mathbf{n}}a_{i}\psi_{i}(u,v) . 
\end{equation*}

\end{enumerate}

\item \emph{Synchronous Updating.} The second deviation rule is where countries deviate simultaneously. We have stipulated that each country assumes others' choices remain the same when it makes its own deviations. This gives us a definition of  ``anticipated linking matrix'' below.

\begin{enumerate} 

\item {\it Anticipated Linking Matrix By Each Country.}  Let $L_{v^{i}_{u}} \in \mathcal{L}_{u}^{i}$ represent \emph{the new linking matrix anticipated by $i$ when deviating from $L_{u}$}, where \emph{only} its $i$-th row is the same as that of $L_{v}$, and the rest of the $n-1$ rows are the same as those in $L_{u}$.  Intuitively,

\begin{enumerate}
\item If $L_{v^{i}_{u}} \neq L_{v}$, the change from $L_{u}$ to $L_{v}$ depends on country $i$ changing the previous linking strategy.  
\item Otherwise, it does not.
\end{enumerate} 

\item {\it Linking Matrix Transition Probability.} The transition probability from $L_{u}$ to $L_{v}$, $\zeta_{uv}$, is the product of every country's individual transition probability from $L_{u}$ to its anticipated linking matrix $L_{v^{i}_{u}}$,

\begin{equation*}
\zeta_{uv} =  \prod\limits_{i \in \mathbf{n}} \psi_{i}(u, v^{i}_{u}).
\end{equation*}

\end{enumerate}

\end{enumerate}

\subsubsection{Further Assumptions}  A country's transition probability between linking strategies can be assumed to satisfy the following basic assumptions: 

\begin{enumerate}
\item We normalize the total probability of staying at original linking matrices and transitioning to different ones in either deviation to be 1.  

In asynchronous updating,
\begin{equation*}
\sum\limits_{L_{v} \in \mathcal{L}_{u}^{i}}\psi_{i}(u,v) = 1.
\end{equation*} 
In synchronous updating, 
\begin{equation*}
\sum\limits_{L_{v^{i}_{u}} \in \mathcal{L}_{u}^{i}}\psi_{i}(u,v_{u}^{i}) = 1
\end{equation*}

\item In either updating, if there does not exist another linking matrix in country $i$'s deviation set mapped to a strictly preferred relation configuration by $\tau$ than the one mapped from $L_{u}$,  we assume the probability of staying at the original linking matrix to be 1. 

If $\not\exists L_{v} \in \mathcal{L}$ such that for country $i$, $\tau(L_{v}) \prec \tau(L_{u})$, then 
\begin{equation*}
\psi_{i}(u,v) = 0, \forall L_{v} \in \mathcal{L}, L_{v} \neq L_{u}, ~\text{and}~ \psi_{i}(u,u) = 1
\end{equation*}

%
%

Otherwise,
\begin{equation*}
\psi_{i}(u,v) > 0.
\end{equation*}

\end{enumerate}

\subsubsection{Deviation Process} The deviation process is formalized using:
\label{process}

\begin{enumerate}\item {\it Initial Distribution.} Countries can be regarded as being initially involved in a distribution of all linking matrices in $\mathcal{L}$.  

Denote the distribution of all the linking matrices in $\mathcal{L}$ at time $T$ as $\pi^{T}$. $\pi^{T} = [\pi^{T}_{u}]^{1 \times |\mathcal{L}|}$,  where $\pi^{T}_{u} \in [0,1]$ representing the probability mass of staying in $u$ at time $T$, and $\|\pi^{T}\|_{1} = 1$. Obviously, $\pi^{0} = [\pi^{0}_{u}]^{1 \times |\mathcal{L}|}$ represents the \emph{initial distribution}.

\item {\it Deviation Process.} The dynamical system is the collection $\Omega = (\Gamma, \pi^{0}, \Zeta\}$. Countries' deviation process can be conveniently understood as applying the operator $\Zeta$ (which is in fact a Markov operator) repeatedly to the initial distribution, the vector $\pi^{0}$. 

The Ces\`{a}ro average as below
\begin{equation*}
\lim\limits_{n\to\infty}\frac{1}{n}\sum\limits_{k=1}^{n}\pi^{0}Z^{k}
\end{equation*} 
represents the limiting distribution of the linking matrices (and equivalently, of the relation configurations) in the deviation process.

\end{enumerate}

%

\subsubsection{Convergence}  In this section we prove in Proposition~\ref{prop/sum} that  $\Zeta$ is always a stochastic matrix. Then we know that a stationary distribution always exists for $\Zeta$, which will give us predictions regarding the long run distribution of linking matrices (or relation configurations) countries involve in.

\begin{prop}
\label{prop/sum} In either asynchronous updating or synchronous updating, every row of $\Zeta$ sums to 1. 

\end{prop}{\bf Proof of Proposition~\ref{prop/sum}:} In asynchronous updating, 

\begin{align*}
\sum\limits_{L_{v} \in \mathcal{L}}\zeta_{uv} = \sum\limits_{L_{v} \in \mathcal{L}}\sum\limits_{i \in \mathbf{n}}a_{i}\psi_{i}(u,v) 
= \sum\limits_{i \in \mathbf{n}}a_{i}\sum\limits_{L_{v} \in \mathcal{L}_{u}^{i}}\psi_{i}(u,v)
\end{align*}

Since $\sum\limits_{L_{v} \in \mathcal{L}_{u}^{i}}\psi_{i}(u,v) = 1$, then we have

\begin{align*}
\sum\limits_{L_{v} \in \mathcal{L}}\zeta_{uv} = \sum\limits_{i \in \mathbf{n}}a_{i} = 1
\end{align*}

In synchronous updating, 

\begin{align*}
\sum\limits_{L_{v} \in \mathcal{L}}\zeta_{uv} = \sum\limits_{L_{v} \in \mathcal{L}}\prod\limits_{i \in \mathbf{n}} \psi_{i}(u,v^{i}_{u}) 
\end{align*}

The above expression can be equivalently expressed as

\begin{align*}
\sum\limits_{L_{v} \in \mathcal{L}}\prod\limits_{i \in \mathbf{n}} \psi_{i}(u,v^{i}_{u}) = \prod\limits_{i \in \mathbf{n}}\sum\limits_{L_{v} \in \mathcal{L}_{u}^{i}} \psi_{i}(u,v) = 1
\end{align*} 

Therefore, under both rules, every row of $\Zeta$ sums to 1. $\square$

We know that an argument for any Markov Chain on finite state space to have a stationary distribution is given by the Brouwer Fixed Point Theorem. Therefore, this dynamical system always has a stationary distribution.

\section{Linkage with Power Allocation Game}
\label{linkage}

Let two power allocation games, which \emph{only} differ in terms of the relation configurations, $R$ and $\hat{R}$, be $\Gamma(R)$ and $\Gamma(\hat{R})$. A reliable basis for a country's preferences for $R$ and $\hat{R}$ is the outcomes from these two power allocation games -- for instance, this country will certainly prefer the relation configuration in which it can survive (i.e., its total support balancing out total threats) in the equilibria of the respective power allocation game over the other in which it cannot. Let their equilibrium sets be $\mathcal{U}$ and $\mathcal{\hat{U}}$. 

Given country $i$, call $\hat{R}$ an ``admissible alternative'' to $R$ if $r_{ik} = \hat{r}_{ik}$ for all $k \in \mathbf{n} \smallsetminus \{j\}$ and $\forall \hat{U} \in \mathcal{\hat{U}}$, $\forall U \in \mathcal{U}$, $x_{k}(\hat{U}) = x_{k}(U)$.

Below we propose a set of the most intuitive and universal criteria regarding countries' preferences for relation configurations, summarized below as ``basic choice axioms''.  

{\it Basic Choice Axioms for $(\mathcal{R}, \preccurlyeq_{i})$.}  
We suppose that country $i$ has $\hat{R}$ and $R$ with the former being an admissible alternative to $R$, and introduce the following two axioms.

\textbf{(1). Comparison along the self relation}

Country $i$ weakly prefers $\hat{R}$ over $R$, $R \preccurlyeq \hat{R}$ if $(\forall \hat{U} \in \mathcal{\hat{U}}$, $\sigma_{i}(\hat{U}) \geq \tau_{i}(\hat{U})) \lor (\forall U \in \mathcal{U}$, $\sigma_{i}(U) \leq \tau_{i}(U))$

\textbf{(2). Comparison along a \text{friend}/\text{adversary} relation}

Country $i$ weakly prefers $\hat{R}$ over $R$, $R \preccurlyeq \hat{R}$ if 
	\begin{enumerate}
	\item
	if $r_{ij} = \hat{r}_{ij} = \text{friend}$, and \\$(\forall \hat{U} \in \mathcal{\hat{U}}, \sigma_{j}(\hat{U}) \geq \tau_{j}(\hat{U}) \lor  (\forall U \in \mathcal{U}, \sigma_{j}(U) \leq\tau_{j}(U)$

	\item
	if $r_{ij} = \hat{r}_{ij} = \text{adversary}$, and  \\$(\forall \hat{U} \in \mathcal{\hat{U}}, \sigma_{j}(\hat{U}) \leq \tau_{j}(\hat{U}) \lor (\forall U \in \mathcal{U}, \sigma_{j}(U) \geq \tau_{j}(U)$

	\item
	if $r_{ij} = \text{friend}$ and $\hat{r}_{ij} = \text{null}$:
		\begin{enumerate}
		\item
		
		$\forall U \in \mathcal{U}$, $\sigma_{j}(U) \leq \tau_{j}(U)$.
		
		\end{enumerate}
		\item
	if $r_{ij} = \text{null}$ and $\hat{r}_{ij} = \text{friend}$:
		\begin{enumerate}
		\item
		
		$\forall \hat{U} \in \mathcal{\hat{U}}$, $\sigma_{j}(\hat{U}) \geq \tau_{j}(\hat{U})$.
		
		\end{enumerate}

	\item
	if $r_{ij} = \text{adversary}$ and $\hat{r}_{ij} = \text{null}$:
		\begin{enumerate}
		\item
		
		$\forall U \in \mathcal{U}$, $\sigma_{j}(U) \geq \tau_{j}(U)$.
		\end{enumerate}
		\item
	if $r_{ij} = \text{null}$ and $\hat{r}_{ij} = \text{adversary}$:
		\begin{enumerate}
		\item
		
		$\forall \hat{U} \in \mathcal{\hat{U}}$, $\sigma_{j}(\hat{U}) \leq \tau_{j}(\hat{U})$.
		\end{enumerate}
	\item
	if $r_{ij} = \text{null}$ and $\hat{r}_{ij} = \text{null}$:
	
	By default, $R \sim \hat{R}$ 
		\end{enumerate}

Now we suppose country $i$ has two arbitrary $\hat{R}$ and $R$, and introduce the last axiom. 

 \textbf{(3) Comparison of the Relation Configurations}

Country $i$ weakly prefers  $\hat{R}$ over $R$, $R \preccurlyeq  \hat{R}$, if 

\begin{enumerate}
\item  all of the respective conditions in (1) and (2) holds for any $j \in \mathbf{n}$.

\end{enumerate}

Due to the linkage with the power allocation game, three features of the \emph{basic choice axioms} are:

First, by capturing \emph{only} the most intuitive aspects in countries' preferences, these axioms only make possible a partial order of the elements in $\mathcal{R}$. However, a valid total order that satisfy these axioms must exist, because the axioms speak to mutually exclusive conditions.

Second, the axioms have considered the important fact in reality that even when changing only a relation, a country may cause elsewhere in the whole environment to change as well, e.g., by influencing the power allocation outcomes of other neighboring or even non-neighboring countries. 


Third, despite not having directly specified, for instance, how a country assesses the relation configurations in terms of the relations among other countries, the axioms did have suggested one way of doing so -- these other countries' relations among themselves could indirectly influence this country's preferences for the relation configurations by directly influencing the power allocation outcomes of its neighbors.

%
%
%
%
%
%

\begin{prop}
\label{prop/rfe}
Assume the basic choice axioms, if  $\exists i \in \mathbf{n}$ such that $p_{i} > \sum\limits_{j \in \mathbf{n} \smallsetminus \{i\}}p_{j}$, which means that a ``superpower'' does exist, a linking matrix $L$ mapped to an optimal relation configuration of $i$ by $\tau$, $\tilde{R}$, is a Nash equilibrium of $\Sigma = \{\mathcal{C}, \mathcal{L}, \tau, \mathcal{R},\preccurlyeq\}$.

\end{prop}

{\bf Proof of Proposition~\ref{prop/rfe}:} When a superpower does exist, i.e., $\exists i \in \mathbf{n}$, $p_{i} > \sum\limits_{j \in \mathbf{n} \smallsetminus \{i\}}p_{j}$,

Suppose one maximal element of the totally ordered set of relation configurations for $i$, which is one of its \emph{optimal relation configuration}, is $\tilde{R}$. 

Note that $\tilde{R}$ must satisfy several properties. 

\begin{enumerate}
\item $\forall j \in \mathbf{n} \smallsetminus \{i\}$, $\tilde{r}_{ij} \neq \text{null}$. According to the basic choice axioms, a \text{null} relation would be always weakly less preferred than a friend relation with the friend being safe or precarious or an adversary relation with the adversary being unsafe or precarious. 

\item $\forall j,k \in \mathbf{n} \smallsetminus \{i\}$, as mentioned earlier, $\tilde{r}_{jk}$ only matters to $i$'s preferences if it affects the neighbors of $i$, all else being equal. 

\end{enumerate}

Correspondingly, $L$ satisfies the following.

\begin{enumerate}

\item first, let $L$ be the linking matrix, by which $i$ determines the relations with all the other countries $\forall j \in \mathbf{n}$, $l_{jj} =  \text{self}$ and $l_{ij} = l_{ji} = \tilde{r}_{ij}$ accordingly to its optimal relation configuration. 

\item second, there are many possibilities for the relations among countries other than $i$ depending on $i$'s optimal relation configuration. For any $j,k \in \mathbf{n} \smallsetminus \{i\}$, let $l_{jk} = l_{kj}$ if $\tilde{r}_{jk} =\tilde{r}_{kj} = \text{adversary}$,  $l_{jk} = l_{kj}$ if $\tilde{r}_{jk} =\tilde{r}_{kj} = \text{friend}$, and $l_{jk} = l_{kj} = \text{null}$ if $\tilde{r}_{jk} =\tilde{r}_{kj} = \text{null}$. 

\end{enumerate}

 $i$ determines the states of all the other countries in the equilibria of $\Sigma$.  In any equilibrium, any friend of $i$ will always be safe or precarious, and any adversary of $i$ will be unsafe or precarious. 
 
None of the countries in $\mathbf{n} \smallsetminus \{i\}$ can deviate from $L$. It cannot unilaterally deviate from the adversary relations, or change the \text{null} relation to a friend relation. And it has no incentives to change a \text{friend} relation into a null relation. 

When a country other than $i$ weakly prefers an adversary relation over a \text{friend} or \text{null} relation, it must be that this new adversary of itself is also an adversary of $i$ but this country itself is not $i$'s adversary. Therefore, this change does not affect $i$'s power allocation outcomes. Therefore, $L$ still realizes an optimal relation configuration for $i$.

%


As argued above, none of the countries have incentives to deviate, including $i$.  Therefore, $L$ is also a Nash equilirbium of $\Sigma$. $\square$

 \section{Application}
 \label{application}
 
 \subsection{Alliance Politics: Cohesion and Effectiveness}
 
An important prediction from the combination of both games is that internal cohesion of countries in alliances is the precondition for achieving any common goal with power allocation. This is an issue usually absent from the formal literature of alliances~\citep{fordham2014all,hiller2011alliance,smith1995alliance,snyder1997alliance,leeds2003alliance,morrow1991alliances,walt2000alliances,olson1966economic}). For instance,  Olson and Zeckhauser (1966) only applies to a narrow subset of alliances that have resolved the internal cohesion issue, such as NATO--the fact that NATO is largely free of the issue of internal cohesion does not mean \emph{every} alliance would function equally well. A combination of the static game of countries' relation formation with the game of countries' power allocation provides some insights into how certain powerful alliances can be both cohesive and effective.

We  present a discussion of ``strong and bonded clique'' that has achieved both internal cohesion and effectiveness, which actually provides a theoretical metaphor for NATO. When this clique exists, we will prove in Theorem 1 that it is really ``reducible'' into a single entity that sums up the total power of all its component countries. By ``reducibility'', we mean that for countries outside of the clique, the clique allocates power \emph{indistinguishably} from this single entity. 

A \emph{strong and bonded clique} can be defined as following: A set of countries $\mathcal{S} \subseteq \mathbf{n}$ is a \emph{strong and bonded clique} if \begin{enumerate}
\item it is a ``clique'' in the sense that any two countries in $\mathcal{S}$ are friends, $\forall i, j \in \mathcal{S}$, $r_{ij} = \text{friend}$;
\item it is strong in the sense that their total power exceeds that of the rest of the countries, $\Sigma_{i \in \mathcal{S}}p_{i} > \Sigma_{j \in \mathbf{n} \smallsetminus \mathcal{S}}p_{j}$; 
\item it is united by bonds -- every country in $\mathcal{S}$ shares the same optimal relation configurations set $\mathcal{\tilde{R}}$, where every element in $\mathcal{\tilde{R}}$ has them in the clique $\mathcal{S}$. 
\end{enumerate}

\subsubsection{A Game's Reduced Form}\label{indis}
For two power allocation games $\Gamma = \{\mathcal{C},  p, \mathcal{U}, \sigma_{i}, \tau_{i}, \mathcal{X}, \preccurlyeq\}$ and $\tilde{\Gamma} =  \{\tilde{\mathcal{C}},  \tilde{p},  \tilde{\mathcal{U}},  \sigma_{i}, \tau_{i}, \tilde{\mathcal{X}}, \preccurlyeq\}$, if there exists a surjection $f: \mathbf{n} \to \tilde{\mathbf{n}}$ such that the following holds
\begin{enumerate}
\item
Clique Members' External Relations Condition: $\forall i, j, h\in \mathbf{n}$ s.t. $f(i) = f(j) \neq f(h)$, $r_{ih} = r_{jh}$. 
\item
Equilibrium and State Condition: $\forall k,l \in \tilde{\mathbf{n}}$, for every equilibrium $U$ of $\Gamma$, let $\tilde{u}_{kl} = \sum\limits_{i \in f^{-1}(k)}\sum\limits_{i \in f^{-1}(l)}u_{ij}$, $\tilde{U}$ is an equilibrium of $\tilde{\Gamma}$. And $\forall i,j \in \mathbf{n}$ s.t. $f(i) = f(j)$, $x_{i}(U) = x_{j}(U) =\tilde{x}_{f_{i}}(\tilde{U}) = \tilde{x}_{f_{i}}(\tilde{U})$. 

\item Total Power Condition: $\forall k \in \tilde{\mathbf{n}}$, $\sum\limits_{i \in f^{-1}(k)}p_{i} = \tilde{p}_{k}$
\end{enumerate}
$\tilde{\Gamma}$ is called a \emph{reduced form} of $\Gamma$ under $f$, especially in the sense that one (or more) subset(s) of vertices in $\mathbf{n}$ can be reduced into a single node, with the total power attached to this single node being the sum of the power attached to the vertices in the subset.

\subsubsection{Theorem 1: Reducibility}
\label{theorem/red}
For two power allocation games $\Gamma = \{\mathcal{C},  p, \mathcal{U}, \sigma_{i}, \tau_{i}, \mathcal{X}, \preccurlyeq\}$ with a strong and bonded clique $\mathcal{S}$ and $\tilde{\Gamma} = \{\mathcal{C} \smallsetminus \mathcal{S} \cup \{s\},  \tilde{p}, \mathcal{\tilde{U}}, \sigma_{i}, \tau_{i}, \mathcal{\tilde{S}}, \preccurlyeq\}$, $\tilde{\Gamma}$ is a \emph{reduced form} of $\Gamma$ under $f: \mathbf{n} \to \mathbf{\tilde{n}}$ if
\begin{enumerate}
\item $\forall i \in \mathcal{S}$, $f(i) = s$.
\item $\forall j \in \mathbf{n}\smallsetminus \mathcal{S}$, $f(j) = j$.
\item $\forall i, j \in \mathcal{S}$, $k \in \mathbf{n}\smallsetminus \mathcal{S}$, $r_{ik} = r_{jk} = \tilde{r}_{sk}$.
\item $\sum_{i \in \mathcal{S}}p_{i} = p_{s}$.

\end{enumerate}

\subsection{Optimal Network Design: Static and Dynamical}

In this section we show how a strong and bonded clique can both induce a relation configuration optimal for itself in the equilibrium of the static game, or gradually improve the odds of being in optimal relation configurations in the dynamical system, respectively in Proposition~\ref{fe} and Proposition~\ref{pi}.
 
\begin{prop}
\label{fe}
In game $\Sigma = \{ \mathcal{C}, \mathcal{L}, \tau, \mathcal{R}, \preccurlyeq\}$ where the parameters for the existence of a strong and bonded clique $\mathcal{S}$ are satisfied and the basic choice axioms are assumed to hold for countries' preferences, an equilibrium linking strategy profile $L$ mapped to an optimal relation configuration $R$ by $\tau$ for countries in $\mathcal{S}$, i.e., \emph{the clique members}, exists.

\end{prop}

\begin{prop}
\label{pi}
In the dynamical system $\Omega = (\Gamma, \pi^{0}, \Zeta\}$, if there exists a strong and bonded clique $\mathcal{S} \subseteq \mathbf{n}$, the probability of playing the set of linking strategy profiles $\mathcal{\hat{L}}$ mapped to the optimal relation configuration set $\mathcal{\hat{R}}$ by $\tau$ for countries in $\mathcal{S}$, $\sum_{L_{p}: ~\tau(L_{p}) \in \mathcal{\hat{R}}}\pi^{T}_{p}$ weakly increase with $T$, under the two deviation rules. 
\end{prop}

This ``optimal network design'' thesis speaks particularly to the literature of unipolar politics \citep{ikenberry2011international,wohlforth1999stability,monteiro2014theory,ikenberry2009unipolarity,monteiro2011unrest} on why the hegemon's (the US's) leader status remains largely challenged.

Quite contrary to the literature which explains the status quo as the product of the power preponderance of the US, we suggest that the status quo may be instead \emph{by design} by the hegemon and certain other great powers, such as by a strong and bonded clique.  In particular, the bonds are both \emph{necessary} and \emph{sufficient} for the \emph{identity} of the clique and that with the theory projected onto the reality, \emph{liberal} bonds hold the clique together in the current world\citep{moravcsik1992liberalism,moravcsik1997taking,oneal1997classical,oneal1996liberal,doyle2005three}. More precisely, the strong and bonded clique actually abstracts for NATO headed by the US in control of the contemporary world.

\section{Conclusion}

To summarize, in terms of the two models of the framework of countries' relation formation, the static game is the simplest possible unifying model of countries' relation formation, while the second model of the dynamical system really develops an view of the long-run processes of countries' relation formation from the perspective of an outsider or an  ``impartial spectator''. 

For future work, we will continue to explore more issues of interest by combining countries' power allocation and relation formation, such as regarding alliances with different levels of cohesion and effectiveness, and a hierarchy of controls or influences different kinds of powerful entities can exert over other countries.

\begin{ack}

We thank Ji Liu for very insightful comments on preparing for the drafts of this paper.  
\end{ack}

%

\bibliography{alliance}

\begin{thebibliography}{22}
\providecommand{\natexlab}[1]{#1}
\providecommand{\url}[1]{\texttt{#1}}
\providecommand{\urlprefix}{URL }
\expandafter\ifx\csname urlstyle\endcsname\relax
  \providecommand{\doi}[1]{doi:\discretionary{}{}{}#1}\else
  \providecommand{\doi}{doi:\discretionary{}{}{}\begingroup
  \urlstyle{rm}\Url}\fi

\bibitem[{Bloch(2009)}]{bloch2009endogenous}
Bloch, F. (2009).
\newblock Endogenous formation of alliances in conflicts.

\bibitem[{Doyle(2005)}]{doyle2005three}
Doyle, M.W. (2005).
\newblock Three pillars of the liberal peace.
\newblock \emph{American Political Science Review}, 99(03), 463--466.

\bibitem[{Fordham and Poast(2014)}]{fordham2014all}
Fordham, B. and Poast, P. (2014).
\newblock All alliances are multilateral rethinking alliance formation.
\newblock \emph{Journal of Conflict Resolution}, 0022002714553108.

\bibitem[{Hiller(2011)}]{hiller2011alliance}
Hiller, T. (2011).
\newblock Alliance formation and coercion in networks.

\bibitem[{Ikenberry et~al.(2009)Ikenberry, Mastanduno, and
  Wohlforth}]{ikenberry2009unipolarity}
Ikenberry, G.J., Mastanduno, M., and Wohlforth, W.C. (2009).
\newblock Unipolarity, state behavior, and systemic consequences.
\newblock \emph{World Politics}, 61(01), 1--27.

\bibitem[{Ikenberry et~al.(2011)Ikenberry, Mastanduno, and
  Wohlforth}]{ikenberry2011international}
Ikenberry, G.J., Mastanduno, M., and Wohlforth, W.C. (2011).
\newblock \emph{International relations theory and the consequences of
  unipolarity}, volume~61.
\newblock Cambridge University Press.

\bibitem[{Leeds(2003{\natexlab{a}})}]{leeds2003alliance}
Leeds, B.A. (2003{\natexlab{a}}).
\newblock Alliance reliability in times of war: Explaining state decisions to
  violate treaties.
\newblock \emph{International Organization}, 57(04), 801--827.

\bibitem[{Leeds(2003{\natexlab{b}})}]{leeds2003alliances}
Leeds, B.A. (2003{\natexlab{b}}).
\newblock Do alliances deter aggression? the influence of military alliances on
  the initiation of militarized interstate disputes.
\newblock \emph{American Journal of Political Science}, 47(3), 427--439.

\bibitem[{Li(2017)}]{yukedis}
Li, Y. (2017).
\newblock \emph{A Network Approach to International Relations}.
\newblock Ph.D. thesis, Yale University.

\bibitem[{Li and Morse(2017)}]{allocation}
Li, Y. and Morse, A. (2017).
\newblock Game of power allocation on networks.
\newblock \emph{Proceedings of American Control Conference}.

\bibitem[{Monteiro(2011)}]{monteiro2011unrest}
Monteiro, N.P. (2011).
\newblock Unrest assured: Why unipolarity is not peaceful.

\bibitem[{Monteiro(2014)}]{monteiro2014theory}
Monteiro, N.P. (2014).
\newblock \emph{Theory of Unipolar Politics}.
\newblock 132. Cambridge University Press.

\bibitem[{Moravcsik(1992)}]{moravcsik1992liberalism}
Moravcsik, A. (1992).
\newblock \emph{Liberalism and international relations theory}.
\newblock 92. Center for International Affairs, Harvard University Cambridge,
  MA.

\bibitem[{Moravcsik(1997)}]{moravcsik1997taking}
Moravcsik, A. (1997).
\newblock Taking preferences seriously: A liberal theory of international
  politics.
\newblock \emph{International organization}, 51(04), 513--553.

\bibitem[{Morrow(1991)}]{morrow1991alliances}
Morrow, J.D. (1991).
\newblock Alliances and asymmetry: An alternative to the capability aggregation
  model of alliances.
\newblock \emph{American Journal of Political Science}, 904--933.

\bibitem[{Olson and Zeckhauser(1966)}]{olson1966economic}
Olson, M. and Zeckhauser, R. (1966).
\newblock An economic theory of alliances.
\newblock \emph{The Review of Economics and Statistics}, 48(3), 266--279.

\bibitem[{Oneal et~al.(1996)Oneal, Oneal, Maoz, and Russett}]{oneal1996liberal}
Oneal, J.R., Oneal, F.H., Maoz, Z., and Russett, B. (1996).
\newblock The liberal peace: Interdependence, democracy, and international
  conflict, 1950-85.
\newblock \emph{Journal of Peace Research}, 33(1), 11--28.

\bibitem[{Oneal and Russet(1997)}]{oneal1997classical}
Oneal, J.R. and Russet, B.M. (1997).
\newblock The classical liberals were right: Democracy, interdependence, and
  conflict, 1950--1985.
\newblock \emph{International Studies Quarterly}, 41(2), 267--294.

\bibitem[{Smith(1995)}]{smith1995alliance}
Smith, A. (1995).
\newblock Alliance formation and war.
\newblock \emph{International Studies Quarterly}, 405--425.

\bibitem[{Snyder(1997)}]{snyder1997alliance}
Snyder, G.H. (1997).
\newblock \emph{Alliance politics}.
\newblock Cornell University Press.

\bibitem[{Walt(2000)}]{walt2000alliances}
Walt, S.M. (2000).
\newblock Alliances: balancing and bandwagoning.
\newblock \emph{International Politics: Enduring Concepts and Contemporary
  Issues}, 96--103.

\bibitem[{Wohlforth(1999)}]{wohlforth1999stability}
Wohlforth, W.C. (1999).
\newblock The stability of a unipolar world.
\newblock \emph{International security}, 24(1), 5--41.

\end{thebibliography}
\bibliographystyle{ifacconf}

\end{document}